\documentclass[aps,prl,twocolumn,superscriptaddress,showpacs,a4paper]{revtex4}

\usepackage{graphicx,amsmath,SIunits}

\usepackage{graphicx,amsmath,SIunits}
\usepackage[latin1]{inputenc}
\usepackage{latexsym,stmaryrd}

\begin{document}
\title{Photonic Crystals with Controlled Disorder}

\author{P.D. Garc\'{i}a \dag}
\author{R. Sapienza \ddag}
\affiliation{Instituto de Ciencia de Materiales de Madrid (CSIC)
 and Unidad Asociada CSIC-UVigo, Cantoblanco 28049, Madrid
Espa\~{n}a.}

\author{C. Toninelli}
\affiliation{European Laboratory for Nonlinear Spectroscopy \&
CNR-INO, 50019 Sesto Fiorentino (Florence), Italy}

\author{C. L\'{o}pez}
\affiliation{Instituto de Ciencia de Materiales de Madrid (CSIC)
 and Unidad Asociada CSIC-UVigo, Cantoblanco 28049, Madrid
Espa\~{n}a.}

\author{D.S. Wiersma}
\affiliation{European Laboratory for Nonlinear Spectroscopy \&
CNR-INO, 50019 Sesto Fiorentino (Florence), Italy}

\date{\today}

\begin{abstract}
Photonic crystals are extremely sensitive to structural disorder
even to the point of completely losing their functionalities.
While on one side this can be detrimental for applications in
traditional optical devices, on the other side, it gives also rise
to very interesting new physics and maybe even new applications.\
We propose a route to introduce disorder in photonic crystals in a
controlled way, by creating a certain percentage of vacancies in
the lattice. We show how the method works and what type of
materials can be obtained this way. Also we characterize the
resulting transport properties from various points of view,
including measurements of the transport and scattering mean free
path and the diffusion constant.
\end{abstract}

 \pacs{42.25.Dd, 42.25.Bs, 42.25.Fx,  06.30.Gv, 66.30.h}

\maketitle

\section{I. INTRODUCTION}

Numerous applications of photonic crystals \cite{Yablo,John} have
been proposed, based on their ability to control ballistic light
transport.\ In practice, most photonic crystals contain a certain
amount of intrinsic disorder, which gives rise to multiple light
scattering and light diffusion \cite{Koenderink}.\ An
extraordinary progress has been made in the fabrication of
nano-photonic structures, with many novel optical properties
\cite{Review}.\ Using engineered disorder by introducing defects
of controlled amount, position, shape, and other morphological
characteristics can, however, lead to interesting new
functionalities \cite{Florendefects,Noda}.\ The extreme case of
disorder is that of a photonic glass, in which the building blocks
are perfect spheres which are distributed randomly \cite{PG2}.\ In
such systems, light undergoes multiple resonant Mie-scattering due
to the equal shape and size of its scatterers \cite{Sapienza}.

While light propagation in photonic crystals is described by
Bloch-modes, transport in photonic glasses is dominated by random
multiple scattering.\ Random systems also exhibit interference
effects, of which maybe the most dramatic is that of Anderson
localization of light waves, in which interference brings light
transport to a complete halt \cite{Anderson}.\ The combination of
a photonic crystal and a certain amount of random multiple
scattering is believed to be the key to observe optical Anderson
localization \cite{John}: the presence of a photonic bandstructure
can lead to a strongly reduced diffusion constant compared to an
equivalent random system without the underlying periodic backbone
\cite{Toninelli}.\ While Anderson localization requires very
strongly scattering materials, the interplay between order and
disorder is mostly unknown, even at low refractive index, far away
from the predicted Anderson localization transition,

In this paper we show a method to obtain structures in between a
photonic crystal and a photonic glass, by adding a small amount of
controlled defects, and to characterize the topology of the
resulting optical properties.\ The paper is organized as follows:
In section II, we will show how to introduce extrinsic disorder in
photonic crystals as vacancies in the lattice and how to analyze
the topological disorder of such structures.\ In section III we
will show measurements of the static transport properties, in
particular the scattering mean free path.\ In section IV we will
report on measurements of the diffusion constant in photonic
crystals as a function of the density of vacancies.\ In section V,
we will show total white light transmission measurements from
photonic glasses and photonic crystals with a high amount of
vacancies.

\section{II. VACANCY-DOPED PHOTONIC CRYSTALS}

\subsection{Sample preparation}

\begin{figure}
    \begin{center}
   \includegraphics[width=8cm]{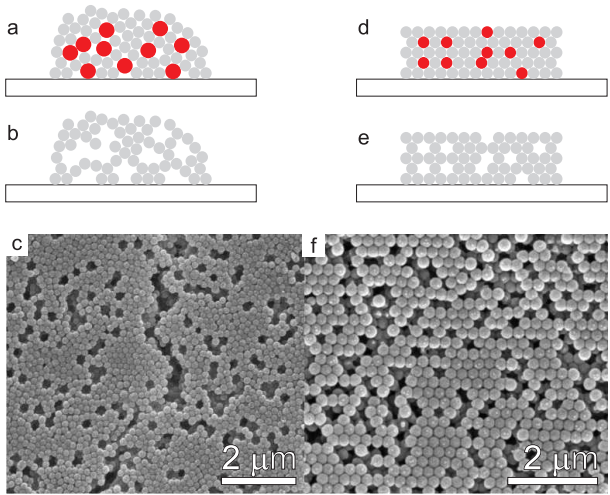}
    \caption{ \label{1} (Color online) \textbf{(a)} The diagram
schematizes the film growth by vertical deposition when
polystyrene spheres (PS, red) and polymethyl-methacrylate spheres
(PMMA, grey) are mixed together in the initial colloidal
suspension with different diameters.\ The structure that forms
depends on the diameter ratio of the spheres and is generally
random.\ By etching with Cyclohexane the PS spheres can be removed
\textbf{(b)}.\ In \textbf{(c)} a SEM image is shown of such
structure.\ \textbf{(d)}, \textbf{(e)} and \textbf{(f)} schematize
the same process when both PS and PMMA spheres have the same
diameter.\ In that case a regular lattice is formed with a
controlled number of vacancies.\ In \textbf{(f)} a sample is shown
that was grown by using PS and PMMA spheres with diameters
$237\,\text{nm}$.}
    \end{center}
    \end{figure}

An \textit{alloy photonic crystal} \cite{BBlaader,BOzin,BSpin} is
grown with a binary colloid that consists of spheres of two
types.\ Compared to the crystals which are composed by
single-specie spheres, binary crystals exhibit a rather rich phase
behavior that depends on the volume fractions of the constituents,
in particular on their diameter ratio.\ If the constituents are
chemically different but of the same diameter, it is possible to
obtain also a regular lattice of which, after crystal growth, one
of the constituents can be chemically removed.\ The one
constituent that is removed acts in that sense as a dopant, since
it introduces vacancies in the lattice.\
Fig.~\ref{1}\textbf{a}-\textbf{b} and \ref{1}\textbf{d}-\textbf{e}
schematize the process.\ Binary colloidal suspensions of
polymethyl-methacrylate (PMMA) and polystyrene (PS) spheres were
ordered by vertical deposition \cite{Vertical}.\ This method
allows us to grow large homogenous 3D alloys with appreciable
thickness.\ The total colloidal concentration in the liquid is
typically set fixed at $0.15\,\text{wt}\%$, and the density of
dopants (PS spheres, in this case) is tuned by changing the
partial PS concentration in the initial colloidal suspension.\ For
example, if a $10\%$ final vacancy density is needed, a partial PS
spheres concentration of $0.015\,\text{wt}\%$ and PMMA
concentration of $0.135\,\text{wt}\%$ will be mixed to obtain a
total $0.15\,\text{wt}\%$ colloidal suspension. Once the alloy
colloidal crystal is grown, the PS spheres are removed by
selective chemical etching, by immersing the samples in $99\%$
pure Cyclohexane for, at least, 4 h.\ This very easy procedure
completely removes the PS spheres, leaving the PMMA spheres
undisturbed.\ Scanning electron microscopy (SEM) image
Fig.~\ref{1}\textbf{f} shows a colloidal crystal realized with
$30\%$ of PS spheres and $70\%$ of PMMA spheres after PS etching.\
In that case, both PMMA and PS spheres had the same diameter $d =
237\,\text{nm}$, with an error included in the polidispersity of
each of the spheres ($<2\%$ of the sphere diameter).

\begin{figure}
    \begin{center}
   \includegraphics[width=8cm]{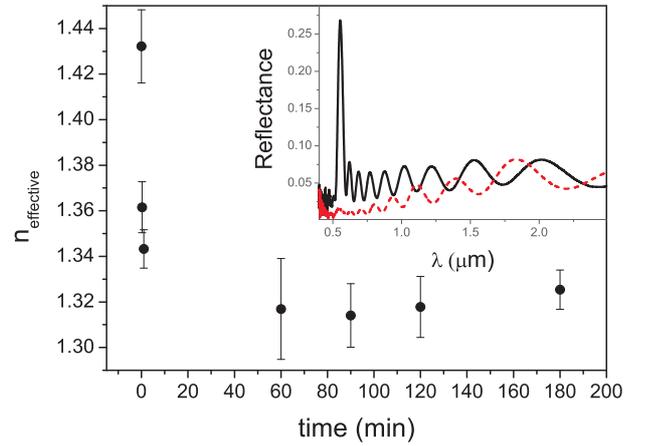}
    \caption{ \label{2} (Color online) Effective refractive index
($n_\text{eff}$) of the system as a function of PS etching time\
The concentration of PS and PMMA spheres are known and the
thickness of the sample, \textit{L}, can be then calculated from
the Fabry-Perot fringes in the reflectance spectrum (black-solid
curve in the inset of the figure)\ By dissolving the PS spheres,
$n_\text{eff}$ reduces and its variation is accounted for by the
blueshift in the Fabry-Perot fringes\ The evolution of
$n_\text{eff}$ can be estimated with the help of \textit{L} and
the Fabry-Perot fringes of the etched sample (red-dashed curve in
the inset of the figure)\ The PS removal has a characteristic time
of about few minutes for which $n_\text{eff}$ reaches a constant
value. }
    \end{center}
    \end{figure}

The selective etching of PS spheres can be conveniently monitored
optically, during the etching process.\ At low energies, far from
the first stopband, the sample can be considered as a homogenous
thin layer with effective refractive index $n_\text{eff}$.\ This
$n_\text{eff}$ can be measured by recording the spectral
separation of the Fabry-Perot fringes in the transmittance or
reflectance spectra, which depend on the {\em optical} thickness
of the sample and thereby both on its physical thickness and
refractive index.\ Local reflectance maxima of the Fabry-Perot
fringes will appear at (for an opal on a substrate with a
refractive index higher than that of the opal \cite{Born}):

\begin{equation}
\label{fabry}
 m \lambda_{m} = 2Ln_\text{eff} \Longrightarrow \frac{1}{\lambda_{m}}=\frac{m}{2Ln_\text{eff}}
\end{equation}
where \textit{m} is the resonance order, \textit{L} is the sample
thickness and $\lambda_{m}$ is the wavelength of the
\textit{m}-resonance.\ A linear relation is obtained by plotting
the inverse of $\lambda_{m}$ as a function of \textit{m}, which
slope yields the inverse of the sample thickness and
$n_\text{eff}$.\ The initial $n_\text{eff}$ of the alloy crystal,
composed by PS and PMMA spheres (Fig.~\ref{1}\textbf{d}), can be
calculated from the partial concentration of each type of spheres,
their refractive index: $n_\text{PS}=1.59$ and
$n_\text{PMMA}=1.4$, and the total filling fraction of spheres in
the crystal $f=0.74$.\ This gives us a precise estimation of the
sample thickness, which remains constant during the etching
process.\ The evolution of the etching can then be follow by
monitoring the value of $n_\text{eff}$ with time, using
Eq.~\ref{fabry} with this initial value of \textit{L}.\
Fig.~\ref{2} shows the effective refractive index for a thin film
initially composed of $60\%$ of PMMA spheres and a $40\%$ of PS
spheres, both of diameter $d = 237\,\text{nm}$.\ The figure
reveals that the etching is complete after few minutes, when
$n_\text{eff}$ reaches a constant value.\ The inset of
Fig.~\ref{2} shows the reflectance spectrum from the composite
before (black-solid curve) and after (red-dashed curve) complete
PS removal.\ Apart from the disappearance of the reflectance peak
related to the first stopband in the $\Gamma\,\text{L}$ direction
(which will be the object of attention in the following section),
a blue shift in the Fabry-Perot fringes clearly shows the
significant change in $n_\text{eff}$.\ The final vacancy-doped
photonic crystals (Fig.~\ref{1}\textbf{f}) have typically the same
dimensions as a thin photonic crystal film, that is, they have a
thickness of several microns and can be several centimeters wide.\
The vacancies are distributed homogeneously throughout the
structure.

\subsection{Optical characterization}
\label{optical}

\begin{figure}
    \begin{center}
   \includegraphics[width=8cm]{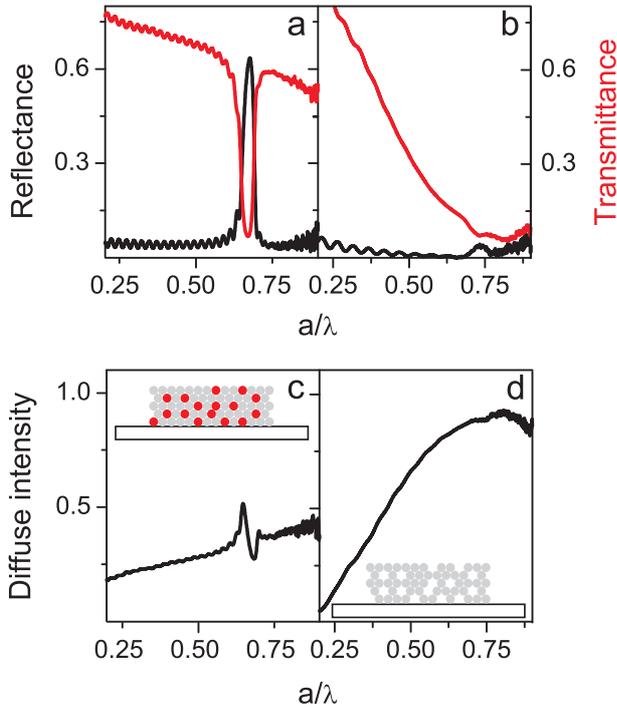}
    \caption{ \label{3} (Color online) \textbf{(a)} and \textbf{(b)}
figures show the reflectance and transmittance spectra measured
from samples with $0\%$ and $40\%$ vacancy density, respectively.\
Diffuse light intensity from the corresponding samples before
\textbf{(c)} and after \textbf{(d)} PS removal.\ Notice the
presence (disappearance) of the gap before (after) PS spheres
removal.\ The insets in \textbf{(c)} and \textbf{(d)} show a
schematical drawing of the corresponding samples. }
    \end{center}
    \end{figure}

As a first characterization of our samples we have recorded
standard angular dependent reflection and transmission spectra.\
In particular, we recorded the specular reflectance, $R$, and
transmittance, $T$, of light with wavelength from $400\,\text{nm}$
to $3\,\micro \text{m}$ in the direction perpendicular to the
sample surface ($\Gamma\,\text{L}$) and in the low energy range
($a/\lambda < 1$, where $a$ is the lattice parameter).\ This
measurement directly gives the amount of light lost from this
particular direction which, due to elastic scattering, propagates
in other directions different than the incident one.\ By
increasing the amount of vacancies in the crystal lattice, the
amount of light scattering increases as we will discuss later.

The analysis carried out in the present section will tackle the
measurement of the amount of remaining order in the bulk of the
structure.\ The magnitude $R+T$ can be used to estimate the amount
of remaining order \cite{GalisteoPRB,AstratovPRB} if absorption
and higher-order Bragg scattering can be neglected.\ In that case,
the diffuse light intensity $D$ is simply $D = 1-R-T$, which
accounts for the losses due to elastic light scattering from the
lattice vacancies which act as the scatterers \cite{note}.\ Note
that one expects the diffuse intensity to be non-zero also for
samples without intentionally introduced defects (our samples with
zero percent vacancies), due to residual polidispersity and
various sample imperfections that are common for even the most
accurately prepared photonic crystal opals \cite{Koenderink}.

\begin{figure}
    \begin{center}
   \includegraphics[width=8.5cm]{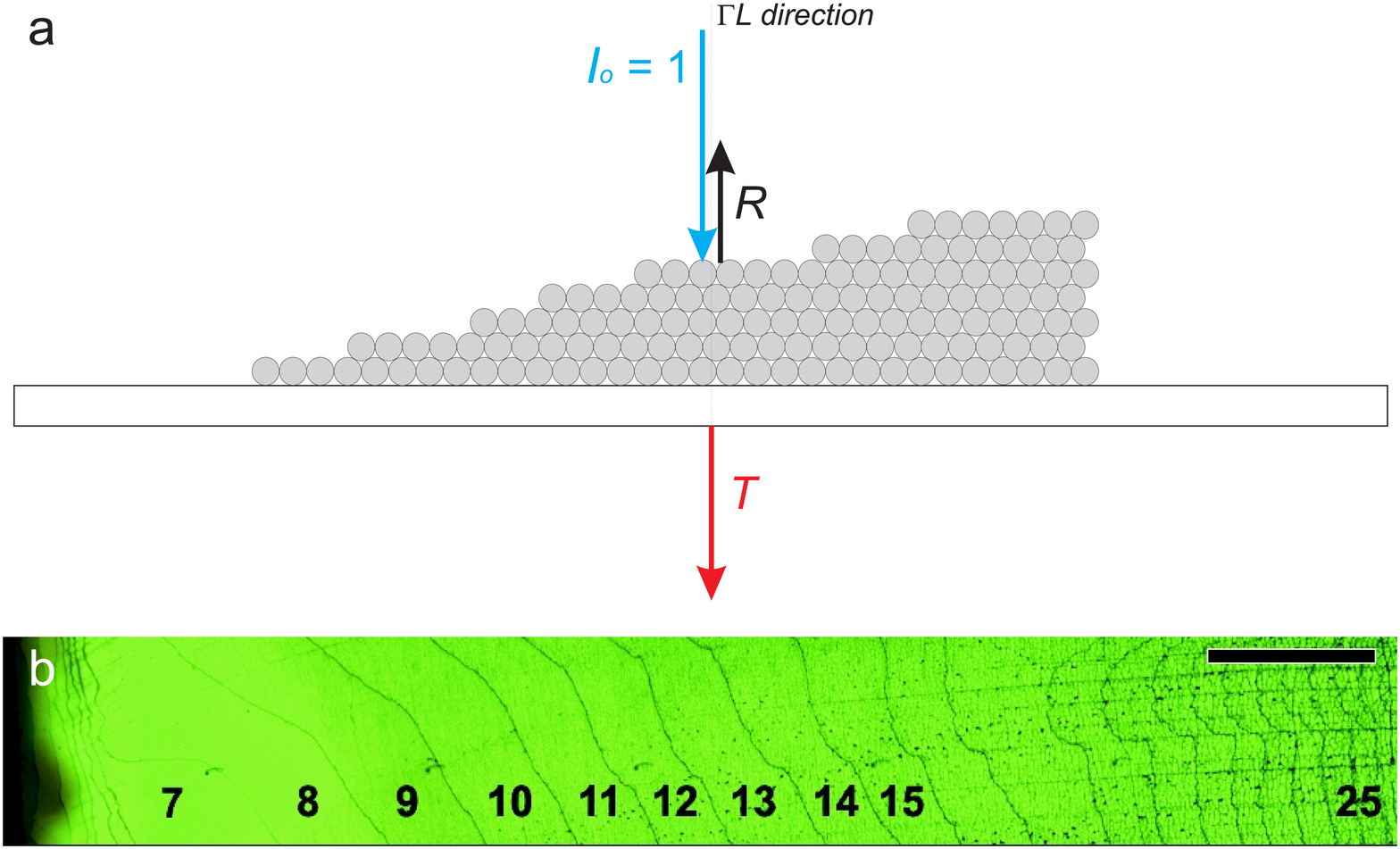}
    \caption{ \label{4} (Color online) \textbf{(a)} Scheme of the
scattering mean free path measurement setup.\ Specular reflectance
and transmittance are measured along the $\Gamma\,\text{L}$
direction (perpendicularly to the sample surface) in adjacent
terraces with known thickness. \textbf{(b)} Optical image made
combining 8 images of the opal surface from a microscope, in which
the different layers are distinguishable as terraces.\ The
measurements have been performed from 1 to 25 layers, which are
clearly distinguishable from each other by eye inspection with the
help of the microscope, on adjacent areas, along horizontal lines
perpendicular to the crystal growth direction. }
    \end{center}
    \end{figure}

Fig.~\ref{3}\textbf{a} and \ref{3}\textbf{b} show $R$ and $T$
measurements from samples doped with a $0\%$ and $40\%$ of
vacancies, respectively.\ The disappearance of the reflectance
(transmittance) peak (dip) related to the pseudogap in the
$\Gamma\,\text{L}$ direction is the first and most evident effect
of increased disorder.\ Fig.~\ref{3}\textbf{c} plots \textit{D},
proportional to the light loses, which presents the usual features
for a finite opal \cite{GalisteoPRB}: a monotonic increase for
frequencies outside the gap, attributed to Rayleigh-Gans type of
scattering, the presence of a dip at $a/\lambda \sim 0.6$ for
polymeric spheres which accounts for a reduction of scattering
losses inside the gap and two peaks at the band edges which
represents an increase of scattering at these spectral positions.\
Here we are measuring light which is lost from the initial
direction and diffuses through the crystal.\ Therefore, the
asymmetry between the high and the low energy band edges can be
explained by examining the photonic band structure in other
crystallographic directions close to the incident one.\ For light
incident in the $\Gamma\,\text{L}$ direction, scattered photons in
the low energy band-edge find allowed states in adjacent
directions when a small momentum is acquired.\ At the high energy
band-edge, the additional momentum needed for a scattered photon
to couple to a Bloch mode in other directions is larger and the
process for this frequency becomes less probable than the former.\
This is the reason which, qualitatively, explains this asymmetry.\
For a high amount of vacancies, Fig.~\ref{3}\textbf{d} reveals the
disappearance of any hint of gap or band-edges features and shows
the increase of scattering losses for all frequencies.\ The higher
the relative concentration of PS, the more abrupt the transition
is: when increasing disorder, constructive and destructive
interference is averaged out giving rise to an effective medium
behavior and losses increase monotonically with energy.

\section{III. Static measurements: Scattering mean free path}
\label{meanvac}

\begin{figure}
    \begin{center}
   \includegraphics[width=8.5cm]{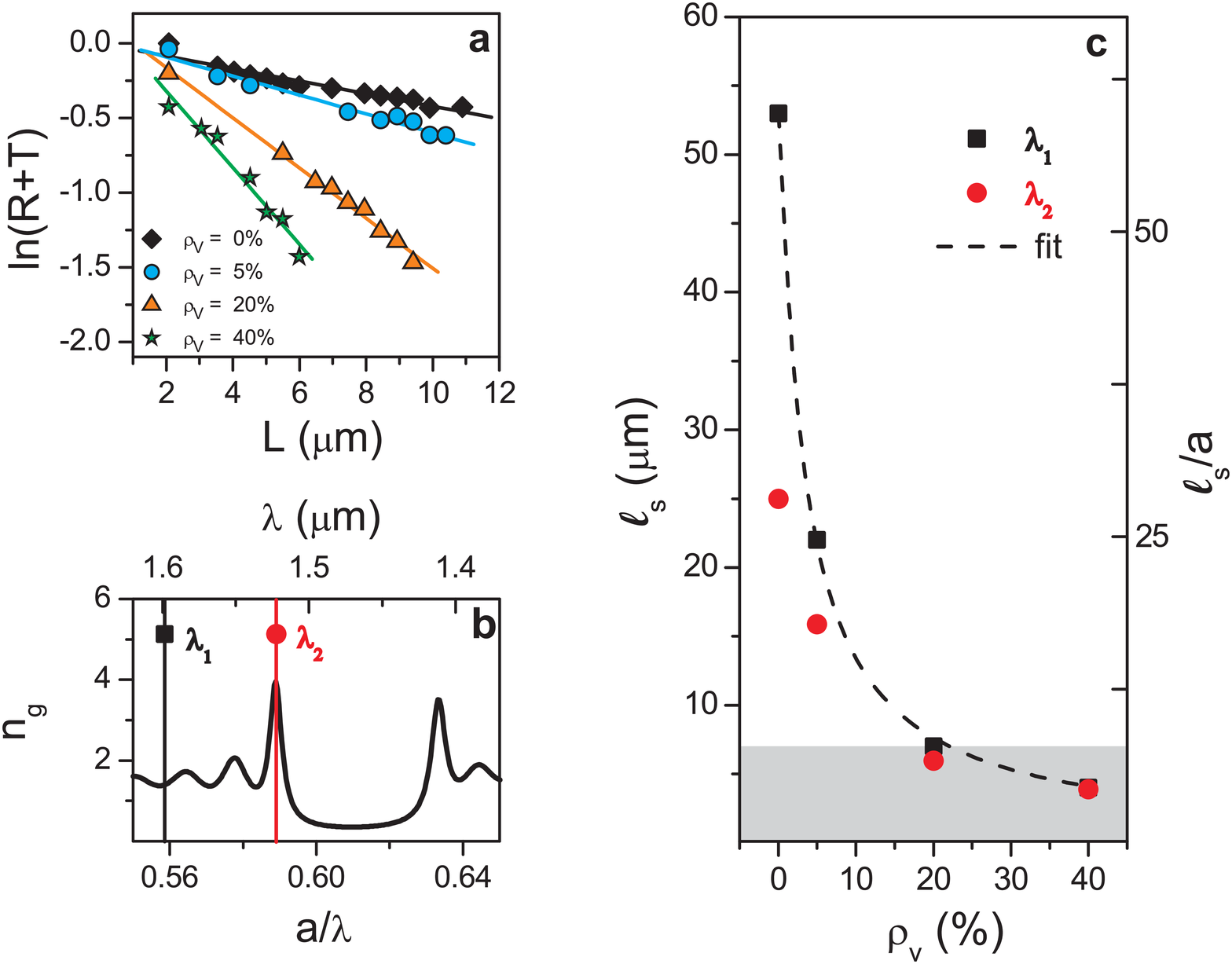}
    \caption{ \label{5} (Color online) Measurement of the scattering mean
free path in photonic crystals.\ \textbf{(a)} Plot of $\mathrm{ln}
(R+T)$ as a function of the sample thickness, $L$, at $\lambda_1
=1.67\,\micro\text{m}$, for different vacancy density (from $0\%$
to $40\%$).\ These photonic crystals are made of PMMA spheres of
630 nm diameter and refractive index $n = 1.4$.\ The inverse of
the slope yields directly the scattering mean free path.
\textbf{(b)} Calculated group index, $n_\text{g}$, for an ideal
photonic crystal of 40 layers thickness along the
$\Gamma\,\text{L}$ direction for the case of $\rho_\text{v} =
0\%$.\ The scattering mean free path, $\ell_\text{s}$, is measured
in the pass band ($a/\lambda_1 = 0.56$, black squares) and at the
band-edge ($a/\lambda_2 = 0.59$, red circles) as a function of the
vacancy density, $\rho_\text{v}$.\ In \textbf{(c)}, the scattering
mean free path is plotted versus vacancy density and compared to
the Bragg length ($L_\text{B}$, shaded area) in the case of
$\rho_\text{v} = 0\%$.\ The variation of $\ell_\text{s}$ is
smaller at the band-edge than in the pass band.\ The black-dashed
curve represents the fit of $\ell_\text{s}$ versus $\rho_\text{v}$
at the pass band. }
    \end{center}
    \end{figure}

The measurements of the reflectance and transmittance coefficient
as reported in the previous section can be used also to determine
the scattering mean free path $\ell_\text{s}$, defined as the
length over which a light beam can propagate inside the sample
before it is scattered due to randomness \cite{PRBscattering}.\ To
that end the reflectance and transmittance are recorded in the way
described in section \ref{optical}, now studied versus the
thickness of the sample (see Fig.~\ref{4}).\ Specular $R$ and $T$
(similar to those from Fig.~\ref{3}\textbf{a} and \textbf{b}) are
taken in adjacent regions in a set of samples with a vacancy
density from $0\,\%$ to $40\%$.\ Adjacent regions have a different
amount of stacked layers and are visible by optical microscope
inspection as terraces on the sample surface.\ The thickness of
such films is assessed by measuring the density of Fabry-Perot
fringes, which is crucial to provide the exact thickness, $L$, of
the tested region.\ The thickness can also be determined in an
alternative and independent way by simply counting the terraces on
the photonic crystal, since each terrace corresponds to a
thickness increase by one layer (see Fig.~\ref{4}\textbf{b}).\ The
accuracy in the determination of the thickness is in that case
limited by the cumulative effect of the sphere polydispersity
($<2\%$).

The analysis of the thickness dependence of reflectance and
transmittance spectra has to take into account the fact that light
impinging on a photonic crystal can be transmitted, (specularly)
reflected, diffracted, absorbed, or (diffusely) scattered.\
Considering energies below the onset of diffraction ($a/\lambda
\sim 1.12$) \cite{GalisteoPRB}, diffraction can be disregarded.\
Absorption is also negligible for the considered frequencies for
PS and PMMA spheres.\ Elastic scattering is then the only loss
mechanism and the Lambert-Beer law can be written as:

\begin{equation}\label{ModifiedLBChapter9}
  R(L) + T(L) = \exp(-\frac{L}{\ell_s})
\end{equation}

Fig.~\ref{5}\textbf{a} shows the measurement of $
\mathrm{ln}(T+R)$ for four different vacancy density at a
wavelength $\lambda_1 = 1.6\,\micro\text{m}$ for spheres with
diameter $d = 630\,\text{nm}$ ($a/\lambda_1 = 0.56$).\ In this
type of representation, the slope yields directly
$(-\ell_\text{s})^{-1}$ according to equation
\ref{ModifiedLBChapter9}.\ Fig.~\ref{5}\textbf{c} shows the
variation of $\ell_\text{s}$ with the vacancy density in the pass
band ($\lambda_1$, see Fig.~\ref{5}\textbf{b}). We observe how the
optical thickness of the sample, $L/\ell_\text{s}$ (where $L$ is
the sample thickness), increases with $\rho_\text{v}$.\
Table~\ref{table:ls} shows this evolution and also gives the
average number of scattering events light performs before exiting
the sample, given by $N=(L/\ell_\text{s})^2$.\ This number is
below one (0.2) in the case of the most perfect opal-based
photonic crystal which is comparable to other very high quality
opals \cite{Baumberg} and it becomes as large as 28 in the case of
$\rho_\text{v} = 40\%$.\ This last value is comparable to opals
grown by centrifugation \cite{KoenderinkCBS} which show an average
number of scattering events of $\sim 15$.\ The very high quality
of the crystals is of paramount importance to access the different
light transport regimes of our samples: near-single scattering in
the case of non-doped crystals up to diffusion in the highest
doping case of $\rho_\text{v} = 40\%$.

\begin{table}
\caption{Optical thickness} 
\centering 
\begin{tabular}{c c c c c} 
\hline\hline 
 $\rho_\text{v} \ (\%)$ & $L_{B}\  (\micro \text{m})$  & $ \ell_\text{s} \  (\micro \text{m})$ & $L/\ell_\text{s}$ & $N$ \\ [0.5ex] 
\hline 
 0  & $8 \pm 1$ & $53 \pm 4$ & 0.4 & 0.2  \\
 5  & & $22 \pm 2$ & 1 & 1 \\
 20 & & $7.0 \pm 0.7$ & 3 & 9 \\
 40  & & $4.0 \pm 0.4$ & 5.3 & 28 \\ [1ex] 
\hline 
\end{tabular}
\label{table:ls} 
\end{table}

From Figs.~\ref{5}\textbf{b} and \ref{5}\textbf{c} it is clear
that the dependence of the scattering mean free path is very
different at a wavelength at the band-edge ($\lambda_2$ in
Figs.~\ref{5}) than it is at wavelengths far away from the
stopband ($\lambda_1$ in Figs.~\ref{5}).\ At the band-edge the
density of states is high and the group velocity low, which
increases the amount of scattering and hence reduces the
scattering mean free path \cite{Froufe,PRBscattering}.\ This
explains the large difference between $\ell_\text{s}$ at the two
wavelengths, for $\rho_\text{v} = 0\%$.\ At increasing
$\rho_\text{v}$, the effect of the photonic crystal diminishes and
at $\rho_\text{v} = 40\%$ there is no difference in the value of
the scattering mean free path for the two wavelengths.\ At
wavelengths far away from the stopband we do not expect strong
effects from the photonic crystal and the inverse of the
scattering mean free path should simply depend linearly on the
density of scattering elements.\ If we distinguish between the
scattering that is intrinsically present in the photonic crystal
(expressed in terms of a density $\rho_0$ and cross section
$\sigma_0$ that represents the average of all intrinsic
scattering, e.g., due to polidispersity of the spheres, cracks,
staking faults and so on), and the scattering introduced by the
vacancies (with vacancy density $\rho_\text{v}$ and vacancy
scattering cross section $\sigma_\text{v}$), we can write the
inverse of the scattering mean free path as:
\begin{equation}
\label{lsv}
 \ell_\text{s}^{-1} = \rho_0 \sigma_0 + \rho_\text{v} \sigma_\text{v}
\end{equation}
assuming independent scattering from the vacancies. Eq.~\ref{lsv}
allows to fit $\ell_\text{s}(\rho_\text{v})$ and extract the value
$\sigma_\text{v} = (0.057 \pm 0.002)\,\micro\text{m}^2$.\ Note
that at both wavelengths $\lambda_1$ and $\lambda_2$ the mean free
path becomes smaller than the Bragg length for $\rho_\text{v}
> 20\%$ (shaded area in Fig.~\ref{5}\textbf{c}), which means that
above those vacancy densities the Bloch approximation fails to
give an accurate description of the propagation process.

\section{IV. Dynamic measurements: DIFFUSION CONSTANT}
\label{diffvac}

\begin{figure}
    \begin{center}
   \includegraphics[width=6cm]{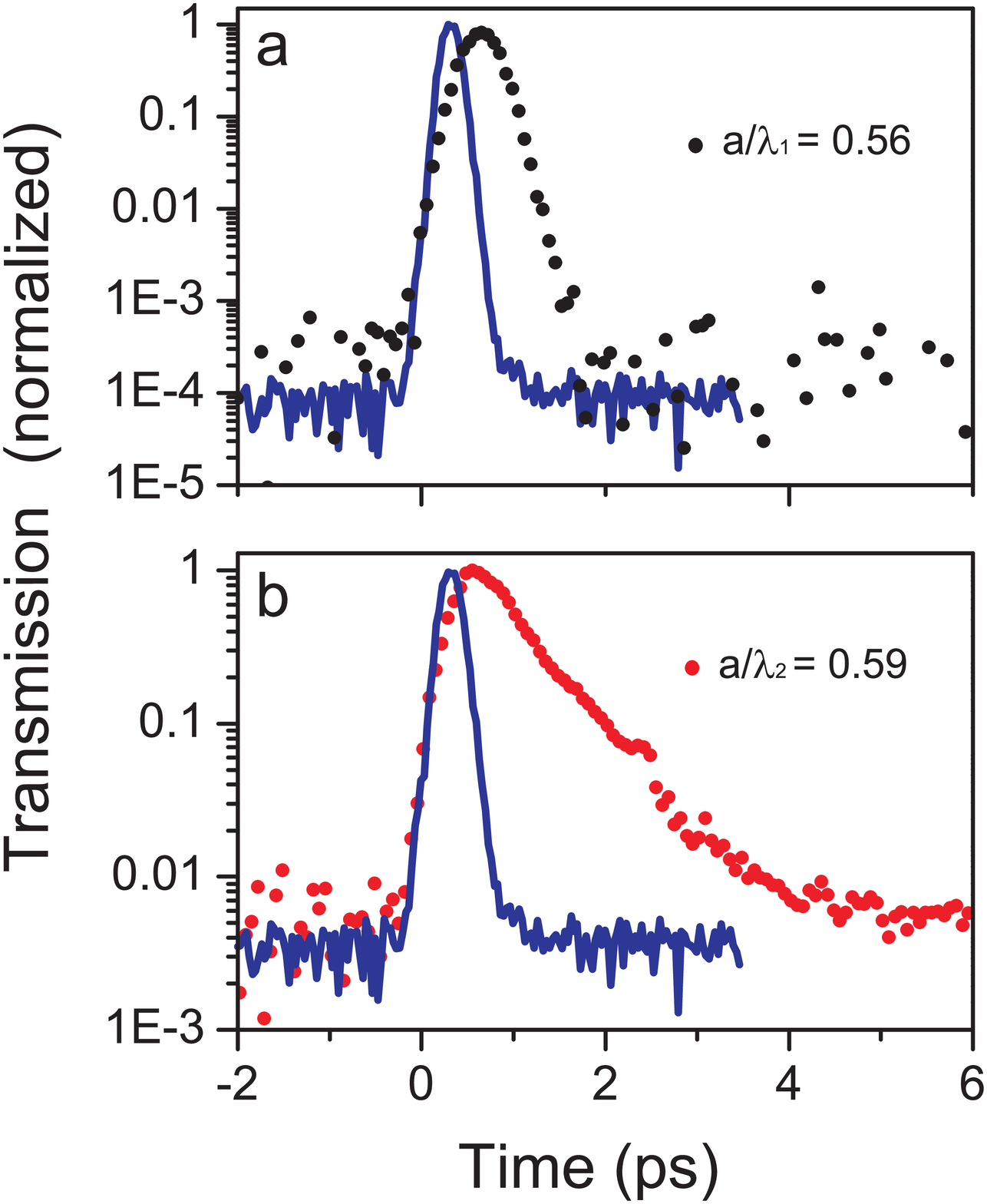}
    \caption{ \label{6} (Color online) Plot of the time-resolved diffuse
transmission through photonic crystals composed by PMMA spheres
with $n = 1.42$ and a diameter $d = 630\,\text{nm}$ with
$\rho_\text{v} = 0\%$ at the pass band, $a/\lambda_1 = 0.56$
black-doted curve \textbf{(a)}, and at the band-edge, $a/\lambda_2
= 0.59$ \textbf{(b)} red-doted curve.\ The blue-solid curve is the
time-resolved transmission of the pulse reference. }
    \end{center}
    \end{figure}

The static measurements presented in section \ref{meanvac} reveal
how sensitive $\ell_\text{s}$ is to the vacancy density.\ In order
to get information on the behavior of the diffusion constant one
has to resort to time-resolved measurements.\ In this section, we
will use a non-linear optical gating technique to analyze the
time-resolved response of transmitted diffuse light through
photonic crystals with vacancies \cite{gating}.\ This will allow
us to measure the diffusion constant as a function of wavelength
and disorder.

Most regular disordered systems are isotropic, meaning that the
diffusion constant and mean free path are angular independent.\ In
(partially disordered) photonic crystals it is, on the contrary,
crucial to take into account directionality.\ The photonic
band-edge of a stopband, for instance, occurs for a wavelength
which will change when varying angle, hence its effect on the
diffuse transport of light can also be anisotropic.\ The technique
that we use in this section to measure the time-evolution of the
transmitted diffuse light is sensitive to the component of the
diffusion constant in the direction perpendicular to the slab,
which is also the direction in which $\ell_\text{s}$ has been
determined in the static measurements described in the previous
section.

Fig.~\ref{6} shows time-resolved transmission through thin film
opals, thickness $21\,\micro\text{m}$ ($\sim 40$ layers), grown
with PMMA spheres (diameter $d = 630\,\text{nm}$) for
$\rho_\text{v} = 0\%$.\ The measurements have been performed at
$a/\lambda_1 = 0.56$ (black-dotted curve) and $a/\lambda_2 = 0.59$
(red-dotted curve).\ The reference pulse is plotted with a
blue-solid curve in both figures to compare it with the diffuse
decay.\ The value of the diffusion constant is obtained by fitting
the experimental time profile of $T(t)$ with the dynamical
solution to the diffusion equation \cite{ltcolloidal1}.\ At the
band-edge, we obtain $\mathcal{D} (\lambda_2) =
220\,\text{m}^2/\text{s}$ while in the pass band, at $\lambda_1$,
the transmitted pulsed through the sample is of the order of the
probe pulse (150 fs, blue-solid curve).\ This is due to the fact
that $\mathcal{D} (\lambda_1)$ is larger than the maximum
diffusion constant we can measure with our setup
($\mathcal{D}_\text{max} \sim 700\,\text{m}^2/\text{s}$).

\begin{figure}[b!]
    \begin{center}
       \includegraphics[width=8.5cm]{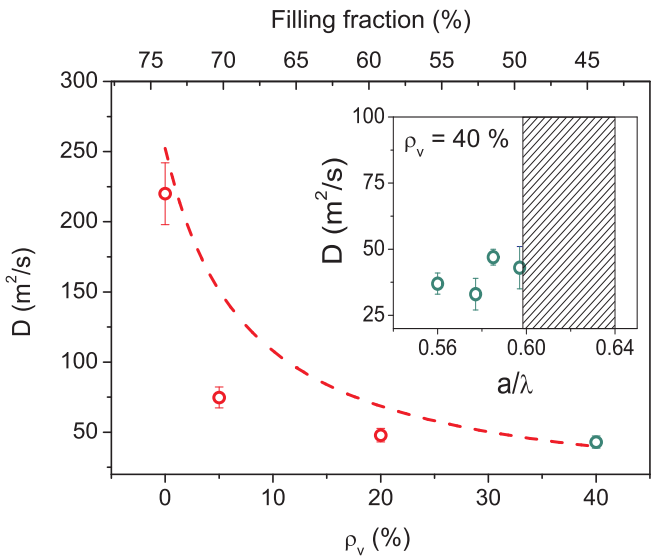}
    \caption{ \label{7} (Color online) Diffusion constant, $\mathcal{D}$,
as a function of the vacancy density, $\rho_\text{v}$, measured at
the band-edge $a/\lambda_2 = 0.59$.\ The red-dashed curve
represents the expected value of $\mathcal{D}$ when assuming a
constant energy velocity and using the vacancy dependence of the
mean free path $\ell_\text{s} (\rho_\text{v})$ as found before.\
One can see that the values at high and low vacancy density are
correctly predicted this way, but the diffusion constant at
intermediate vacancy densities is over-estimated.\ In the inset of
the figure the measured $\mathcal{D}$ is plotted as a function of
energy for $\rho_\text{v} = 40\%$ (the position of the pseudogap
is represented by the dashed area).}
    \end{center}
    \end{figure}

The optical thickness of the sample, $L/\ell_\text{s}$, is
wavelength dependent at $\rho_\text{v} = 0\%$, as shown in the
previous section.\ In the pass band, $\ell_\text{s}(\lambda_1)$
acquires a very large value compared to $L$ and
$\mathcal{D}(\lambda_1)$ becomes very large, as well.\ At this
wavelength, scattering becomes dominantly single/low-order
scattering so that it is actually not meaningful anymore to define
a diffusion constant in the first place.\ The observed strong
wavelength dependence diminishes at increasing disorder.\ This is
shown in the inset of Fig~\ref{7} and is expected, as the photonic
crystal correlations disappear.\ This behavior is clearly observed
in the inset of Fig.~\ref{7}, where $\mathcal{D}$ is plotted at
$\rho_\text{v} = 40\%$.

The complete measurement of $\mathcal{D}(\lambda_2)$ as a function
of $\rho_\text{v}$ is plotted in Fig.~\ref{7}.\ It shows a 5-fold
decrease of the value of $\mathcal{D}$ as a consequence of
disorder from $\mathcal{D}(0\%) = 220\,\text{m}^2/\text{s}$ to
$\mathcal{D}(40\%) = 43\,\text{m}^2/\text{s}$.\ The diffusion
constant already reaches its minimal value at $20\%$ vacancy
density and increasing the vacancy density beyond that point seems
not to reduce the diffusion constant further.\ The total decrease
of $\mathcal{D}$ at $\lambda_2$ is comparable to the decrease of
$\ell_\text{s}$ at the same wavelength.

The diffusion constant in regular isotropic disordered systems is
given by:
\begin{equation}
\label{d}
 \mathcal{D} = \frac{1}{3} \ell_\text{t} v_\text{e}
\end{equation}
where $v_\text{e}$ is the energy velocity \cite{Albada} and
$\ell_\text{t}$ is the transport mean free path.\ This relation
remains valid in isotropic systems for each Cartesian coordinate
if we take the appropriate values for $\mathcal{D}$,
$\ell_\text{t}$, and $v_\text{e} $ in those directions.\ As
mentioned above, our experimental technique is sensitive to the
value of the diffusion constant in the direction perpendicular to
the slab, which is also the direction in which we have determined
the other optical properties of our samples.

If we now assume, to first order, $\ell_\text{t}\approx
\ell_\text{s}$ (which is reasonable for our system) we can use the
values of $\ell_\text{s} (40\%)$ and $\mathcal{D} (40\%)$ to
estimate $v_\text{e} $, which yields: $v_\text{e} (40\%) \sim
0.25\text{c}$ (where c is the speed of light in vacuum).\ This
value is small compared to the phase velocity in an equivalent
homogeneous system with the same average refractive index as our
materials, which lies between 1.3 and 1.4, and also much smaller
then the transport velocity that one would expect for the
equivalent fully disordered system.

In Fig.~\ref{7}, we have also plotted the vacancy density
dependence of the diffusion constant if we assume the transport
velocity to remain constant, using only the vacancy dependence of
the mean free path $\ell_\text{s} (\rho_\text{v})$ as found
before.\ We see that the overall variation of the diffusion
constant of a factor of 5 can be entirely explained in that case
by the vacancy dependence of $\ell_\text{s} (\rho_\text{v})$.\
However, at intermediate vacancy values the such obtained curve
overestimates $\mathcal{D}$.\ Currently there is no theory
available that can describe accurately the behavior of the
transport velocity in partially disordered photonic crystals, and
this would be an interesting topic to look into in the future.

\section{V. TOTAL WHITE LIGHT TRANSMISSION: FROM CRYSTALS TO GLASSES}
In the previous sections we have shown via static and dynamic
measurements how a high amount of vacancies gives rise to strong
multiple scattering in photonic crystals.\ In this section, we
want to compare these highly doped photonic crystals with photonic
glasses.\ To this purpose, we measured total light transmission
from $400\,\text{nm}$ to $900\,\text{nm}$ wavelength through
photonic crystals with $\rho_\text{v} = 40\%$ with an integrating
sphere (the setup is shown elsewhere \cite{PRA}) on large areas of
$\sim \text{mm}^2$.

\begin{figure}[t]
    \begin{center}
   \includegraphics[width=7cm]{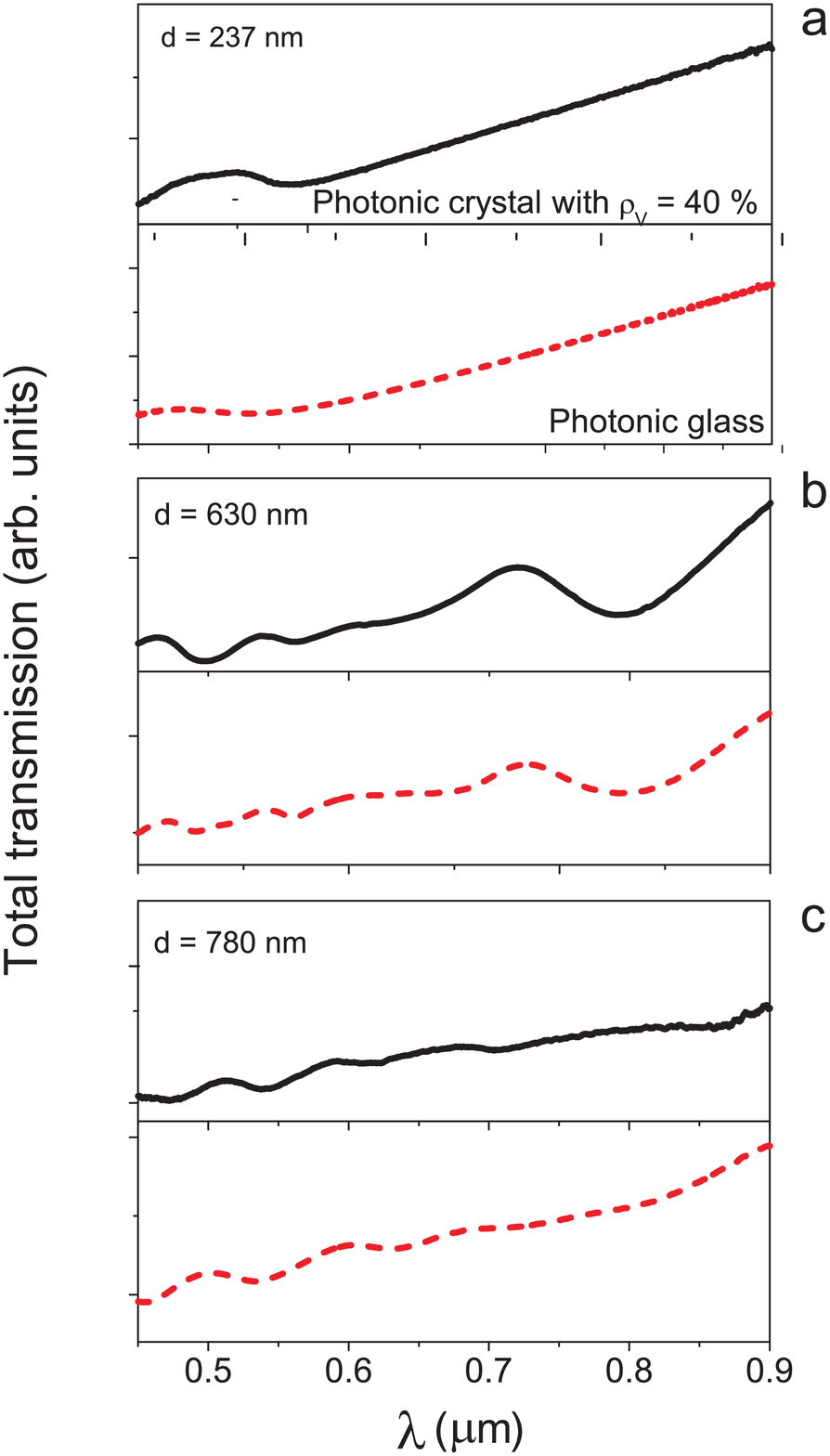}
    \caption{ \label{8} (Color online) Total light transmission through
photonic crystals with $40\%$ vacancies (black-solid curves)
compared to photonic glasses made with the same spheres
(red-dashed curves) with diameters $d = 237\,\text{nm}$
\textbf{(a)}, $d = 630\,\text{nm}$ \textbf{(b)} and $d =
780\,\text{nm}$ \textbf{(c)}.}
    \end{center}
    \end{figure}

Fig.~\ref{8} compares total light transmission through photonic
crystals with $\rho_v=40\%$ vacancies and through photonic glasses
made with the same PMMA spheres for three different diameters:
(\ref{8}\textbf{a}) $d = 237\,\text{nm}$, (\ref{8}\textbf{b}) $d =
630\,\text{nm}$ and (\ref{8}\textbf{c}) $d = 780\,\text{nm}$.\ The
three figures show that the total transmission presents a very
similar behavior in both cases, including the occurrence of very
similar resonances.\ The transport mean free path measured in
photonic glasses, $\ell_\text{t} \simeq 3\,\micro \text{m}$
\cite{Sapienza} is comparable to the one measured in
natural-sedimented photonic crystals \cite{kundePRA} and to the
value of $\ell_\text{s}$ obtained previously for our photonic
crystals with $\rho_\text{v} = 40\%$ vacancies $\ell_\text{s}
\simeq 4\,\micro \text{m}$.\ The occurrence of the same resonances
in the total transmission for photonic glasses and photonic
crystals with vacancies is noteworthy.\ In photonic glasses these
resonances originate from the (Mie) resonances in the single
scattering coefficient of the monodisperse constituent spheres.\
In the doped photonic crystals the scattering originates from
vacancies and apparently, at high vacancy density, the Mie
resonances of the spheres that constitute the photonic crystal
emerge as dominant when the spatial correlations of the photonic
crystal are strongly attenuated. The remarkable similarity of the
curves shown in Fig.~\ref{8} is a clear indication of the
convergence of a photonic crystal into a photonic glass with
disorder.

\section{VI. CONCLUSIONS}

In this paper we report on the realization of photonic crystals
with a controlled amount of, well-defined, disorder.\ We
characterize the optical properties of such materials in various
ways and determine the degree of scattering at various wavelengths
and doping concentrations.\ In particular, our measurements show
that it is possible to control \textit{and fine tune} the amount
of multiple scattering in a photonic crystal, by adding vacancies
and hence \textit{without altering the crystal structure}.\ In the
highly vacancy doped photonic glass we observe a resonant
behavior, analogous to that observed due to Mie scattering in
photonic glasses.\ Our materials might therefore be useful to
explore Fano-like interactions \cite{Fano} between the extended
Bloch-mode of the photonic crystal and spatially confined
Mie-modes.

The possibility of controlling light scattering and diffusion in
photonic crystals has important implications to test and
understand better the quality of photonic crystal-based devices.\
The use of high refractive index materials, like Si
\cite{Ibisate}, may amplify the effect presented here becoming
proper candidates to observe and control Anderson localization of
light in 3D \cite{John83}, as already observed in slightly
disordered 1D photonic crystal waveguides
\cite{Science,waveguides} and for the spectral control of lasing
emission from ordered/disordered active media.

\section{ACKNOWLEDGMENTS}

We thank J. F. Galisteo-Lopez for the data of the group index. The
work was supported by the EU through Network of Excellence
IST-2-511616-NOE (PHOREMOST), and partially supported by EU FP7
NoE Nanophotonics4Energy grant No. 248855; the Spanish MICINN
CSD2007-0046 (Nanolight.es), MAT2009-07841 (GLUSFA) and Comunidad
de Madrid S2009/MAT-1756 (PHAMA) projects . RS acknowledges
support by RyC.



\begin{thebibliography}{99}
\bibitem{note}
This consideration is valid for our samples, since PS and PMMA
absorbance is negligible for the considered frequencies
\cite{PSabsorbance,PMMAabsorbance}, confirmed also by measurements
of the absorption length ($\ell_\text{a} \sim 10\,\text{m}$) in
the visible range for a photonic glass composed by the same
polymer spheres \cite{PRA}. Higher-order scattering is also
negligible compared to the scattering losses.
\end{thebibliography}
\end{document}